\documentclass[a4paper,11pt]{article}
\usepackage[a4paper,includemp=false,body={16cm,24.7cm},vcentering]{geometry}
\usepackage[latin1]{inputenc}
\usepackage[T1]{fontenc}
\usepackage{mathptmx} 
\usepackage[bf,center]{caption}
\usepackage{graphicx} 
\usepackage{wrapfig}

\usepackage{url,amsmath,amssymb,amsthm}

\makeatletter
\renewcommand{\section}{\@startsection{section}{1}{0mm}{30pt}{12pt}{\normalfont\normalsize\bfseries}}

\renewcommand{\subsection}{\@startsection{subsection}{2}{0mm}{18pt}{12pt}{\normalfont\normalsize\itshape}}
\makeatother

\newcommand{\Title}[1]{\begin{center}{\bfseries\fontsize{12pt}{12pt}\selectfont#1}\end{center}}
\newcommand{\Author}[2]{\begin{center}{\fontsize{12pt}{12pt}\selectfont#1}\\{\it #2~}\end{center}}
\newcommand{\Introduction}{\section*{Introduction}}
\newcommand{\Conclusion}{\section*{Conclusion}}

\begin{document}


\Title{Discovery and Characterization of Trans-Neptunian Binaries in Large-Scale Surveys}
  
\Author{Alex H. Parker$^1$}{1. Harvard-Smithsonian Center for Astrophysics, Cambridge MA USA. aparker@cfa.harvard.edu}
 
 \begin{centering}
{\tiny Appears in the proceedings of the \textit{Orbital couples: ``Pas de deux'' in the Solar System and the Milky Way} workshop, Paris, France, October 10---12 2011.}
\end{centering}
 
\Introduction

\noindent

The dynamically cold component of the Kuiper Belt is host to a population of very widely separated, near-equal mass binary systems. Such binaries, representing the tail of the separation distribution of the more common, more tightly-bound systems, are known to have on-sky separations up to $\sim4''$. Their wide separations make them highly valuable due to their delicacy and sensitivity to perturbation, and also makes them relatively easy targets to characterize from the ground. Parker et al. (2011) present a ground-based characterization of seven such systems with separations at discovery ranging from $0.''5 - 4''$, and we will adopt these systems as the prototypes for the ultra-wide binaries of the Kuiper Belt. Here we present the prospects for using future large-scale ground-based optical surveys (with LSST as our baseline survey) to measure the orbital properties of a large sample of these widely separated Trans-Neptunian Binaries (TNBs). 

\vspace{-20pt}
\section{Seeing distribution for LSST observations}

The seeing at the LSST site is well measured, and the optical characteristics of the LSST imager have been extensively modeled. A summary of the estimated delivered seeing is presented in the LSST Science Book (LSST Science Collaboration 2009), and Table 1 contains the FWHM of the delivered PSF in the $r$ and $i$ bands. We adopt these percentiles for determining the repeated resolvability of TNB systems.

\begin{wraptable} {0}{0.6\textwidth}
\vspace{-20pt}
\centering
\begin{tabular}{ ccc }
\multicolumn{3}{c}{\bf Table 1: LSST Site Seeing Distribution}\\

\hline
\multicolumn{3}{c}{Measured $r$ and $i$ seeing (LSST09)} \\
\hline
25th percentile & 50th percentile & 75th percentile \\
$0.''65$ & $0.''76$&$0.''89$ \\

\hline
\multicolumn{3}{c}{Simulated $r$ and $i$ seeing (\S\ref{sims})} \\
\hline
25th percentile & 50th percentile & 75th percentile \\
$0.''64$ & $0.''76$&$0.''89$ \\
\hline
\end{tabular}
\end{wraptable}

The seeing is expected to be better at longer wavelengths ($z$ and $Y$), but we do not consider them here as the limiting magnitude expected per visit in these bands is not as useful for the typical brightness of TNOs.

For simulations in \S\ref{sims}, we require a continuous function PDF for the seeing distribution in order to generate synthetic observational circumstances. We adopt a PDF for the measured seeing distribution at the LSST site which closely reproduces the peak and 25th, 50th, and 75th percentiles of the $r$ and $i$ band (see Table 1). 

\vspace{-20pt}
\section{Projected binary separation}

The fraction of observations which are resolved by LSST will drop for binaries with smaller primary radii, given the same Hill sphere separation distribution as a function of radius. This is because the Hill spheres of TNOs scale linearly with their radii, so a TNB with a smaller primary radius will have a smaller on-sky separation for a physical separation of a given fraction of its Hill sphere. The observed fraction of binaries as a function of primary radius must be corrected for the fraction of a Hill sphere which is resolvable as a function of primary radius in its discovery survey. Such characterization is critical, as the trend of binary fraction with primary radius is a strong indicator of the collisional history of the Kuiper Belt (Petit \& Mousis 2004, Nesvorn\'{y} et al. 2011, Parker \& Kavelaars 2012), providing an orthogonal constraint from size distribution measurements.

Assuming that the binary system is distant and on a circular heliocentric orbit, so $r_\odot \sim r_\oplus$, the projected Hill radius can be estimated for the following two cases:  given the primary and secondary radii $R_P$ and $R_S$, when $R_P \gg R_S$, the projected Hill radius $R_{H} \simeq 18".3 (\rho / \mbox{1 g cm}^3)^{ \frac{1}{3}  } ( R_{P}/ 100 \mbox{km} )$, while in the case that $R_P \simeq R_S$, $R_{H} \simeq 23".1 (\rho / \mbox{1 g cm}^3)^{ \frac{1}{3}  } ( R_{P}/ 100 \mbox{km} )$.

The second case, where the secondary is of comparable mass to the primary, will be used throughout the rest of this paper.

Grundy et al. (2008) found that the most probable projected separation of TNB systems at random epochs (for random orientations and a variety of eccentricity distributions) was within one percent of the mutual semi-major axis ($a_m$), with 84\% of the observed separations being within a factor of two of $a_m$. Therefore for our analytical estimate we adopt $a_m$ as the characteristic separation from which to estimate the on-sky separation.

\begin{wraptable} {0}{0.6\textwidth}
\vspace{-20pt}

\begin{centering}
\begin{tabular}{ lccc }
\multicolumn{4}{c}{\bf Table 2: Radii of resolvable equal-mass TNBs}\\
\hline
$a_m / R_{H}$ & \multicolumn{3}{c}{R$_P$ for which system is resolved in...} \\
\hline
& 25th percentile & 50th percentile & 75th percentile \\
0.05 & 56 km & 67 km & 77 km \\
0.10 & 28 km & 33 km & 36 km \\
0.20 & 14 km & 16 km & 19 km \\
\hline
\end{tabular}
\end{centering}
\vspace{-10pt}
\end{wraptable}

Table 2 demonstrates the implications of the variation of Hill radius size changing with primary radius, illustrating the primary radius at which a binary with a given $a_m / R_H$ must be in order for that separation to translate into an on-sky separation sufficient to be resolved under 25th, 50th, and 75th percentile conditions at LSST. For smaller primary radii, only objects occupying larger fractions of their Hill spheres will be resolved by the survey.

At present six known TNBs fall in the $0.05 \lesssim a_m / R_H \lesssim 0.1$ range (Parker et al. 2011, Shepherd et al. 2011), while just three binaries fall in the $0.1 \lesssim a_m / R_H \lesssim 0.2$ range, and only one TNB is known with $a_m / R_H > 0.2$ (the exceptional system 2001 QW$_{322}$).

\subsection{Separation of known ultra-wide TNBs}\label{sims}	

\begin{wrapfigure}{r}{0.5\textwidth}
\vspace{-65pt}
\begin{center}
\includegraphics[width=0.5\textwidth]{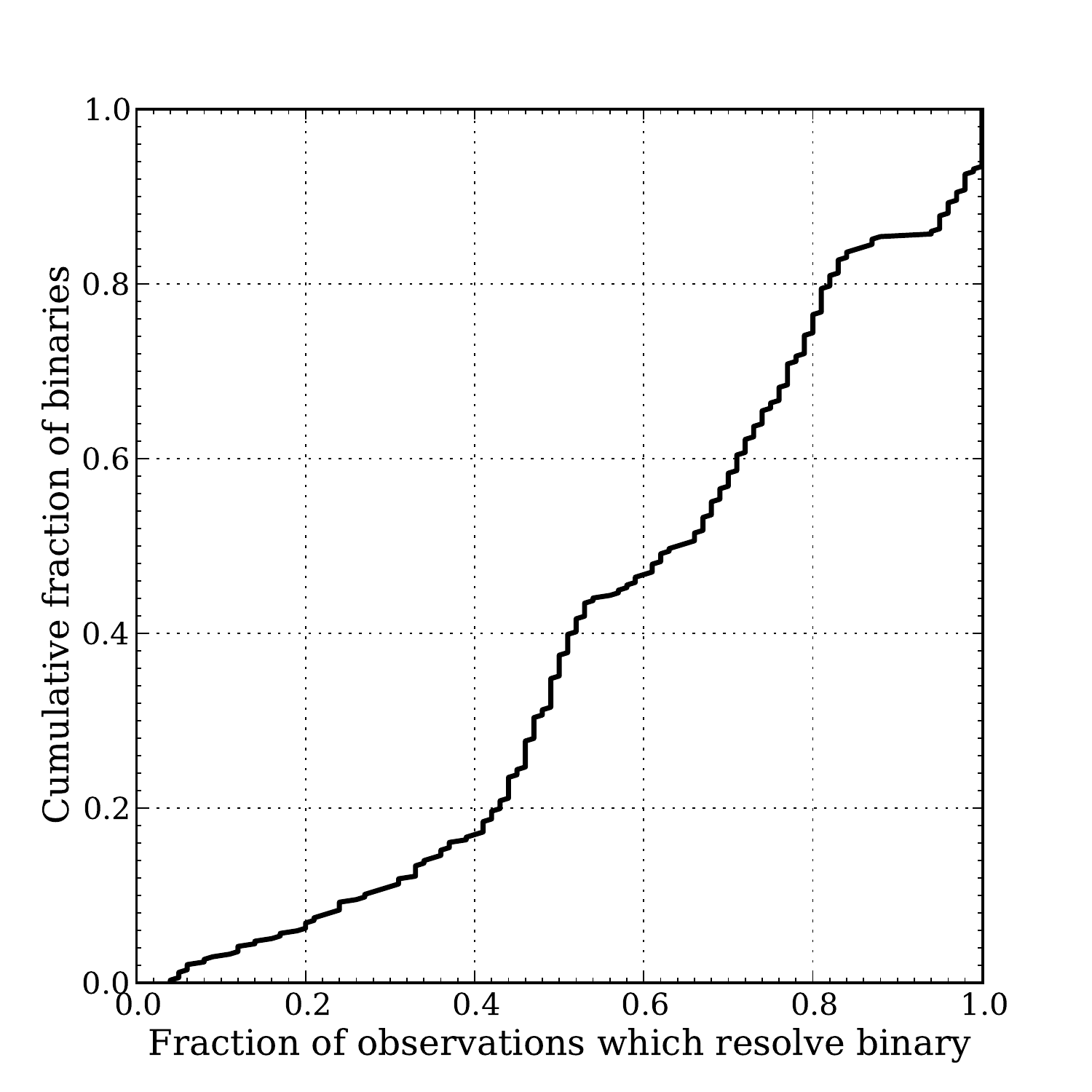}
\caption{Fraction of LSST observations for which binary systems with mutual orbits drawn from Parker et al. (2011) are resolved --- eg., only $\sim5$\% of all binaries with properties comparable to those in Parker et al. (2011) are resolved in fewer than 10\% of visited epochs by LSST --- the remaining 95\% of wide TNBs will be sampled in more than 10\% of the total number of observations.}
\end{center}
\vspace{-30pt}
\end{wrapfigure}

In addition to these analytical estimates, we simulate the observations of known wide TNBs in order to determine the fraction of LSST observations which will resolve systems with comparable properties. The binaries studied in this work are drawn from Parker et al. (2011), which represent the widest binaries with characterized orbits. These seven systems were found to occupy between 7\% and 22\% of their mutual Hill spheres, with physical separations of several tens to one hundred thousand kilometers. Mutual eccentricities varied from $\sim0.2$ to $0.9$. 

Given the baseline of LSST observations of ten years, we sample each system's mutual orbit over ten year segments of that system's heliocentric orbit centered at random epochs. Over these ten year segments, we make synthetic ``observations'' of each system by determining the observed separation of the system at 100 random epochs, given the Earth-binary separation and the angular orientation of the system. This process is repeated for hundreds of random ten-year segments for each binary system.

At each synthetic observation, the LSST seeing distribution was used to generate an estimate of the seeing at the time. If the binary separation at the epoch exceeded the FWHM of the sampled seeing, the binary was said to be resolved during this epoch. Figure 1 illustrates the results of this exercise. We find that all seven of the mutual orbits in Parker et al. (2011) would be resolved in some fraction of observations from the LSST site, only 5\% of all possible orientations of these seven systems will be resolved in fewer than 10\% of all images taken from the LSST site --- with the other 95\% of possible orientations resolvable in more than 10\% of the same images. Given the baseline of 230 $r$-band observations per field over the 10 year LSST survey, this would net $>23$ resolved epochs for 95\% of these ultra-wide TNBs.

\vspace{-20pt}

\section{Estimates of sample size}

The estimated H-magnitude completeness for LSST TNO observations is greater than 90\% completeness (differential) for $H \lesssim 7.5$ (LSST Science Collaborations 2009). The CFEPS survey (Petit et al. 2011) estimate that in the cold classical Kuiper Belt there are approximately 4,000 objects brighter than $H\sim8$, and that the luminosity function of this population is steep with $\alpha \sim -1.2$. This would indicate roughly 1,000 objects brighter than $H\sim7.5$ which would be recovered in the highly complete LSST sample. Extending to the 50\% (differential) completeness magnitude for LSST, $H\sim8.6$, we would have to extrapolate the results of the CFEPS survey over more than half a magnitude. Adopting the same luminosity function slope over this range, we estimate approximately 20,000 TNOs brighter than this magnitude; because of the steepness of the luminosity function, most detections will be in the last few luminosity bins and we can conservatively approximate the differential completeness of 50\% as the cumulative completeness and adopt an estimate of 10,000 TNO detections from the cold classical Kuiper Belt brighter than $H\sim8.6$.

The brighter, more complete sample also conveniently corresponds to a radius of $\sim77$ km (when adopting a geometric albedo of 10\%). This is the radius for which even binaries occupying just 5\% of their mutual Hill sphere will be resolvable in the 75th percentile conditions of the LSST site (Table 2). Given the literature estimate of a wide binary fraction of $\sim 1.5$\% (Kern et al. 2006, Lin et al. 2010), we would expect roughly 15 of the 1,000 TNOs detected by LSST in this bright sub-sample to be resolved binaries. Such binaries, being bright and easily separated, will have high signal-to-noise measurements of their mutual astrometry in many visits, and consequently their mutual orbits will be very well characterized by the LSST observations alone.

The fainter, less complete sample will return far more resolved binaries. The limit of $H\sim8.6$ corresponds to a radius of approximately 46 km. This radius is slightly smaller than the limit resolvable in the 25th percentile seeing at LSST given an $a_m / R_H \lesssim 0.05$. However, this is larger than the limit resolvable in even the 75th percentile seeing for $a_m / R_H \sim 0.1$. As a crude estimate, four of the 10 known wide binaries have separations in excess of 0.1$R_H$, so we can estimate the fraction of binaries in this separation class to be $0.4 \times 0.015 \sim 0.006$. We then expect 60 resolved binaries in this class found by LSST, given our estimate of 10,000 cold Classical TNOs detected in total. Such binaries will have relatively low signal to noise observations as they are near the photometric limit of LSST, but they will be resolved under most conditions and orbital characterization will be possible.

If we consider the limiting primary radius which can be resolved in 25th percentile conditions by LSST given a separation in excess of 0.05$R_H$, we find $R_P \simeq 56$km which translates to $H\sim8.2$ given an albedo of 10\%. This magnitude limit is comparable to the magnitude at which the CFEPS estimates are defined, with approximately 4,000 brighter objects in the cold classical Kuiper Belt  in total. Given a wide binary fraction of 1.5\%, we expect 60 resolved binaries in this sample. While these objects will have moderate signal to noise observations, they will be resolved under a smaller number of circumstances ($\sim25$\%) than the previous two cases. 

The preceding sub-samples have some overlap, and when accounting for the variation of observable fraction of Hill sphere with radius along with the increasing number of TNOs in the sample with decreasing radius, the sum total of TNBs expected to be resolved in at least the 25th percentile conditions (and therefore with $\gtrsim60$ resolved observations in $r$ band alone) is approximately 100. This binary sample represents the systems that will be orbitally characterized by LSST itself; given the rapid increase of binary fraction with decreasing mutual separation (eg., Kern \& Elliot 2006) and that for eccentric systems the observed separation can significantly exceed the projected mutual semi-major axis in some fraction of observations (eg., Grundy et al. 2008), a significantly larger number of systems will likely be discovered by LSST but not directly orbitally characterized. 

\vspace{-20pt}

\section{Value of Large Mutual Orbit Sample}

\begin{itemize}
\item Exceptionally detailed dynamical decomposition: Do prograde TNBs exhibit a marked difference in mutual orbit properties compared to retrograde? Such a difference might indicate separate formation pathways dominating for prograde and retrograde orientations (Schlichting \& Sari 2008, Nesvorn\'y et al. 2010, Parker et al. 2011).
\vspace{-5pt}
\item Detailed host population statistics: Does the Kernel component of the cold Classicals (Petit et al. 2011) or the collisional family of Haumea (Brown et al. 2007) host wide binaries? Are there any such wide binaries amongst the hot Classicals? Is the one resonant wide equal-mass binary (Sheppard et al. 2011) an anomaly, or does it represent a population of resonant wide TNBs?
\vspace{-5pt}
\item Is the fraction of binaries correlated with primary radius, indicating collisional modification of the Kuiper Belt (Nesvorn\'y et al. 2011, Parker \& Kavelaars 2012)?

\end{itemize}

The final point is particularly important, as a correlation of binary fraction with radius would be particularly interesting from a dynamical standpoint; however, it would also significantly undercut the probability of detecting a large sample of wide binaries with LSST. Sheppard et al. (2011) present a survey which discovered 1,000 faint TNOs and only find one system (the brightest in their sample) to be a binary. If such a low wide-binary fraction is characteristic of the small objects in the cold Classical population, then the likely sample size of wide binaries discovered by LSST will be drastically reduced; the estimates of the number of binaries in the bright subset discussed previously would be the least affected. However, the binary 2000 CF$_{105}$ has a primary radius of $\sim 32$ km, making it extremely sensitive to collisional disruption. Its continued existence argues against a strongly collisionally evolved Kuiper Belt (Parker \& Kavelaars 2012) and given that its albedo is high ($\sim0.2-0.3$) relative to the mean of the cold Classicals it may represent a bright member of a large population of small Kuiper Belt binaries.

\vspace{-20pt}

\Conclusion

The population of ultra-wide TNBs are valuable tracers of the dynamical history of the outer Solar System, but at present the conclusions that can be drawn from them are limited by small-number statistics. Barring any strong correlation of binary fraction with primary radius, LSST will characterize roughly 100 ultra-wide TNBs and discover many more. Such a sample will allow substantially stronger conclusions to be made regarding the origins of these systems and their implications for the history of the outer Solar System. Careful debiasing will allow disentangling of albedo distribution and binary fraction trends, further constraining the collisional grinding of the Kuiper Belt.

\end{document}